%% file: main.tex
\documentclass[manuscript]{acmart}
\setcopyright{none}
\renewcommand\footnotetextcopyrightpermission[1]{}
\usepackage{booktabs}
\usepackage{graphicx}
\graphicspath{{figs/}}
\newcommand{\code}[1]{\texttt{\small #1}}

\begin{document}

\title{Only Accepted Tokens Count: An Energy Accounting of Speculative Decoding
Under Latency Targets}

% Authors are suppressed by the `anonymous' class option for review; drop
% that option for the camera-ready version.
\author{Bole Ma}
\orcid{0009-0006-6536-1044}
\affiliation{\institution{Erlangen National High Performance Computing Center}\city{Erlangen}\country{Germany}}
\email{bole.ma@fau.de}
\author{Ayesha Afzal}
\orcid{0000-0002-8632-3681}
\affiliation{\institution{Erlangen National High Performance Computing Center}\city{Erlangen}\country{Germany}}
\email{ayesha.afzal@fau.de}
\author{Jan Eitzinger}
\orcid{0009-0000-3350-3841}
\affiliation{\institution{Erlangen National High Performance Computing Center}\city{Erlangen}\country{Germany}}
\email{jan.eitzinger@fau.de}
\author{Harald Koestler}
\orcid{0000-0002-6992-2690}
\affiliation{\institution{Erlangen National High Performance Computing Center}\city{Erlangen}\country{Germany}}
\email{harald.koestler@fau.de}
\author{Gerhard Wellein}
\orcid{0000-0001-7371-3026}
\affiliation{\institution{Erlangen National High Performance Computing Center}\city{Erlangen}\country{Germany}}
\email{gerhard.wellein@fau.de}
\renewcommand{\shortauthors}{B. Ma et al.}

\begin{abstract}
Large language model serving is limited by facility power. Energy
per token determines how many tokens a datacenter can serve, while a latency
target bounds how quickly each user receives them. During decode, every weight
is read from memory for every generated token. Batching amortizes this read
across users but makes no individual user faster. Speculative decoding does: a
drafter proposes several future tokens, and the target model verifies them in a
single forward pass that reuses the same weight read. Because every drafted
token must be verified whether or not it is accepted, the energy speculation
saves depends on how many drafted tokens are accepted.

This saving is hard to pin down. It varies across target models from different
attention families and across drafting schemes, and it depends on the text
being served. Every drafter we evaluate accepts fewer tokens on conversation
than on code, and every trained drafter accepts fewer on Romansh, a
low-resource language, than on English.

We measure speculative decoding under production serving conditions (SGLang,
CUDA graphs enabled) on H100 and H200 GPUs, at batch sizes up to $64$ across 5 corpora. On dense models, verification, not drafting, dominates the energy cost. The saving shrinks as
the batch grows, because batching has already shared the weight read that
speculation would divide among the accepted tokens. At large batch sizes,
block drafters push verification past the balance point between arithmetic
and memory traffic into the compute-bound regime, where the saving becomes a loss, and a drafter with low
acceptance raises energy per token. Meeting a tighter latency target with
speculation then costs additional energy. Larger batches lower energy per
token without speculation. Speculation adds to that
saving only when acceptance is high, and on the model we tested it leaves
little headroom for latency-preserving DVFS.
\end{abstract}


\maketitle

\section{Introduction}
\label{sec:intro}
Language model serving is limited by facility power. NVIDIA presents its
Vera Rubin systems against a fixed facility power budget and names performance
per watt the defining metric of AI infrastructure~\cite{kampman2026hotchips,nvidia2026perfwatt},
and from the operator side SemiAnalysis reports that OpenAI is limited by
datacenter power rather than by budget or floor space, which makes tokens per
megawatt the quantity it buys~\cite{semianalysis2026jalapeno}. Tokens per
megawatt is energy per token inverted, so every joule a serving configuration
saves per token is capacity a power-limited facility gains~\cite{wattlytics2026}. The same
comparisons already mix systems served with and without speculative
decoding~\cite{semianalysis2026jalapeno}, which is the setting whose energy
this paper measures.

Energy is not the only target. A service is also held to latency objectives
that its users feel: the time to the first token of a response, and the time
between the tokens that follow. A service that saves energy by streaming
tokens more slowly loses its users, so the configuration an operator wants is
one that meets both targets at once. Choosing the lowest-energy configuration
that still meets a latency objective is the framing of cluster-level energy
managers such as DynamoLLM~\cite{DynamoLLM}; this paper asks the same question
of speculative decoding. The first token has several levers, among them prefix
caching~\cite{sglang2024} and running prefill on separate
machines~\cite{patel2023splitwise}. The time between tokens is set by decode
alone.

Decode generates text one token per forward pass. Every pass reads all of the
model's weights to produce a single token, so the machine spends its time
moving bytes while its arithmetic units wait on memory, and both the energy
per token and the time between tokens follow the bytes.

That makes meeting both targets plausible. Reading a byte from off-chip memory costs far more energy than an arithmetic
operation on it: on a GH200, a Hopper GPU like ours, a byte read from HBM costs
about 93\,pJ and a 16-bit tensor-core operation about 0.5\,pJ~\cite{antepara2025benchmark},
so a pass that produces more tokens per weight read should lower energy per
token, and if those tokens belong to one user, such a pass should shorten the time
between that user's tokens as well.

Batching is the first candidate. Running many sequences together reads the
weights once and amortises that read, so energy per token falls and throughput
rises. Batching does nothing for any individual user. On the dense models we measured, the time between a user's tokens grows by 9 to 34\% from a batch of 1 to 32. On the Mixture-of-Experts models it grows 2.5 to 4.0 times,
because each added request routes to the experts chosen for its tokens and adds their weights
to the read.

The time-between-tokens objective is written in that quantity. When the
objective is tight enough, no batch size meets it, and
speculative decoding becomes the only remaining lever. A drafter proposes
several tokens at low cost, the model verifies them in one pass, and the accepted
ones are produced for the cost of a single weight read. On our models speculation shortens the mean time between tokens at single
stream by $1.3$ to $4.3\times$ (Table~\ref{tab:tbt}).
Without speculation, an objective of $25$\,ms per token is missed at every
batch we ran by gemma-4-31B-it, the largest dense model, and an objective of $10$\,ms by 4 of our 11 models; speculation brings every one of them under the
objective.

The remaining question is the energy cost of speculation. Where batching has
already amortised the weight read, there is less for a drafter to recover, and
the drafter still has to be paid for.

This paper measures that trade. We run 5 published drafting schemes
against 11 models spanning 4 attention designs, sweeping the number of
tokens verified per step and the batch. For every model, scheme and batch we
also run a non-speculative baseline, the same model served at the same batch
size without speculative decoding, and we add hardware performance counters for
the roofline and a draft and verify split measured by removing the draft.
The paper contributes (1)~a paired energy measurement of speculative decoding
as production engines serve it; (2)~a mechanism, from the draft and verify
split and from hardware-counter rooflines on two GPUs that show where
verification crosses the balance point and the saving ends; (3)~guidance on the
choice of drafter and the number of drafted tokens across batch and GPU; and
(4)~measurement pitfalls that change the reported saving without any sign in the
result file, among them a CUDA-graph bias of up to 39 percentage points on
Mixture-of-Experts models.

\section{Background: Decode, Speculation, and Draft Cost}
\label{sec:background}
On an H100 the balance point between arithmetic and memory sits near $316$
operations per byte, and on an H200 near $163$, from the tensor-core rates and
memory bandwidths we measure (Section~\ref{sec:measurement}), while a decode
step at single stream executes about 10 operations per byte, including the
padded work of tensor-core tiles (Section~\ref{sec:roofline}).
Decode runs more than an order of magnitude below the balance point, so the
arithmetic units are mostly idle and the energy of a token follows the bytes
that produced it and the time spent moving them.

Speculation raises the arithmetic done per byte by giving each weight read
several tokens of the same request to process. A drafter proposes $k$
tokens and the model being served, the \emph{target model} of the
speculative-decoding literature, checks all of them in one pass, a small
matrix product rather than a vector product. The model keeps the drafted
tokens up to the first one it disagrees with and adds one token that it predicts itself. A
rejected token was still drafted and still carried through verification, so
the drafter's work on that token and its share of the pass return nothing. At
single stream that share costs little, because the extra tokens ride along on
a weight read that dominates the step. At serving batch the pass is already
doing useful work for every request, and the share spent on rejected tokens
displaces useful work. The quantity that
matters is therefore the number of tokens accepted per verification, set
against the cost of every drafted token, and not the fraction of drafted
tokens that is accepted: only the accepted tokens divide the weight read.

We compare 5 drafting schemes: n-gram lookup, multi-token-prediction (MTP)
heads, EAGLE3~\cite{li2025eagle3}, DFlash~\cite{chen2026dflash} and
DSpark~\cite{deepspec2026}. They differ in where the drafter comes from, and
therefore in what a draft costs. Training-free n-gram drafting matches the
recent context against itself and runs no network; SGLang, the engine we
measure, matches on the host~\cite{sglang2024} and vLLM on the
GPU~\cite{vllmngramgpu}, and either way its GPU cost is verifying the tree of
candidates it proposes. The trained drafters come in two kinds. A \emph{draft
head} (the MTP heads, EAGLE3, and gemma's assistant, described below) is one or
a few extra layers that read the hidden state of the model and predict the next
token, so one pass of the head costs a small fraction of a pass of the model.
A head drafts one token per pass and each pass waits for the previous one, so
a draft of $k$ tokens takes $k$ small passes run in sequence. A \emph{block
drafter} (DFlash and DSpark) is a separate small network that drafts a
whole block of tokens in one pass. The block has a fixed size set in training,
counting the last accepted token that opens it: 16 tokens for the z-lab
DFlash drafters, and 7 tokens for DeepSpec's DFlash and DSpark
drafters~\cite{chen2026dflash,deepspec2026}.

A drafter trained together with its model is co-trained, as are the MTP
heads shipped with Qwen3.6, GLM-4.7-Flash and JoyAI-LLM-Flash. A drafter
trained after its model is post-hoc, and its training comes either from the
model's vendor or from a third party. Gemma~4 ships a vendor post-hoc draft
head, trained by Google, which its authors call an assistant~\cite{gemma4};
the engine serves it through its MTP path, so the figures list it under MTP.
The EAGLE3, DFlash and DSpark checkpoints we use are third-party post-hoc
drafters. Every scheme sends the last accepted token through the verification
pass together with the drafted ones, so the model verifies one token more per
request and step than the drafter proposes; for a block drafter this number
equals the block size. Table~\ref{tab:drafters} lists the number of tokens each
drafter has verified per step, and the figures append it to the name of the
drafter: DFlash-16 verifies 16 tokens per request and step.

\section{Measurement}
\label{sec:measurement}
The 11 models span 4 attention families: plain grouped-query attention
(GQA: Qwen3-4B, Qwen3-8B, Qwen3-14B and Qwen3-Coder-30B-A3B-Instruct),
grouped-query attention with sliding-window layers (GQA+SWA: gemma-4-12B-it,
gemma-4-26B-A4B-it and gemma-4-31B-it), multi-head latent attention (MLA:
GLM-4.7-Flash and JoyAI-LLM-Flash), and state-space hybrids that interleave
Gated DeltaNet layers with gated attention (GDN+GA: Qwen3.6-27B and
Qwen3.6-35B-A3B). Of the 11 models, 5 are Mixture-of-Experts models, marked with asterisks in
Table~\ref{tab:drafters}.
Every run uses one NVIDIA H100 SXM5 with 94\,GB of HBM2e and a 700\,W power
limit, on which we measure a memory bandwidth of 2.28\,TB/s and a BF16
tensor-core rate of 719\,TFLOP/s. Of the 11 models, 3 run their batch sweeps on an H200 with 141\,GB of HBM3e
(4.44\,TB/s and 722\,TFLOP/s measured) instead: JoyAI-LLM-Flash, whose
weights alone fill the H100's memory, and the two Qwen3.6 hybrids, which the H100 cannot hold at large batch
(Section~\ref{sec:drafter}); Section~\ref{sec:roofline}
repeats Qwen3-4B on the H200 as well. Every run generates 512 tokens per request: a batch of $B$ is $B$ requests
submitted together and decoded to that length, with no further arrivals. Each
batch is drawn from a pool of 96 distinct prompts for its task and sent through
the chat template of each model with its default settings
(Figure~\ref{fig:prompts}), so Qwen3, Qwen3.6 and GLM-4.7-Flash generate
thinking text, and their results describe that text. The code pool feeds every
batch sweep; Section~\ref{sec:task} repeats the sweeps on conversation and long
documents and serves English and Romansh news at batches of 1 and 16.
\newcommand{\pbox}[3]{\fbox{\begin{minipage}[t][#2][t]{#1}\raggedright\footnotesize #3\end{minipage}}}
\newcommand{\tok}[1]{\textcolor{gray}{\texttt{\detokenize{#1}}}}
\begin{figure*}[t]
\centering
\pbox{0.475\textwidth}{2.6em}{\textbf{Code} (every batch sweep)\hfill 16--18 tokens\\[2pt]
\emph{Write a Python function that reverses a list of integers and returns the
result as a tuple.}}\hfill
\pbox{0.475\textwidth}{2.6em}{\textbf{Conversation}, real user turns (Section~\ref{sec:task})\hfill 16--19 tokens\\[2pt]
\emph{how do I add multiple new columns in m for power query or power bi?}}\\[4pt]
\pbox{0.475\textwidth}{3.9em}{\textbf{Long document} (Section~\ref{sec:task})\hfill 1000--1200 tokens\\[2pt]
\emph{Summarize the following government report in three sentences.
The structure of the armed forces is based on the Total Force concept, which
recognizes that all elements of the structure \ldots}}\hfill
\pbox{0.475\textwidth}{3.9em}{\textbf{News, Romansh} (Section~\ref{sec:task}), same instruction\hfill 230--240 tokens\\[2pt]
\emph{Per reducir il dumber da peschs en il lai da fermada Zervreila èn vegnidas
instradadas differentas mesiras. \ldots}}\\[4pt]
\pbox{0.475\textwidth}{4.8em}{\textbf{News, English} (Section~\ref{sec:task})\hfill 240--280 tokens\\[2pt]
\emph{Continue the following news article. Write the next few paragraphs in the
same language as the article and in the same style. Do not translate, summarise
or comment on it. Steven Gerrard and Phil Jagielka stood shoulder-to-shoulder
\ldots}}\hfill
\pbox{0.475\textwidth}{4.8em}{\textbf{As sent}, the code prompt in each model's chat template\hfill +5--40 tokens\\[2pt]
Qwen3-14B: \tok{<|im_start|>user} \emph{Write a Python function \ldots}
\tok{<|im_end|>} \tok{<|im_start|>assistant}\\[1pt]
gemma-4-12B-it: \tok{<bos><|turn>user} \emph{Write a Python function \ldots}
\tok{<turn|>} \tok{<|turn>model} \tok{<|channel>thought} \tok{<channel|>}}
\caption{One prompt from each pool, with the section that uses it and the range of its median length in tokens across the models' tokenizers, and the code prompt as two models receive it after their chat templates, template tokens in grey.}
\label{fig:prompts}
\end{figure*}
Energy comes from the cumulative energy counter built into the GPU
(\texttt{nvmlDeviceGetTotalEnergyConsumption}), read at $10$\,Hz; the energy of
a generation is the difference between two readings. The counter covers the GPU
board only, not the host processor, host memory, network or cooling, which
favours n-gram drafting, whose matching runs on the host. Weights and the
key-value cache are BF16, decoding is greedy, each cell runs at least 3 seeds,
and each model uses one SGLang build (0.5.12 or newer) under CUDA 13.2. Every speculative run is compared with a
non-speculative baseline run: the same model served without speculative
decoding, at the same batch size, with the same prompt pool, generation length,
GPU and engine settings. The
energy saving is the fraction by which the energy per token of the speculative
run falls below that of this baseline run, so a difference in serving
configuration cannot be reported as a difference in scheme.
Acceptance is ragged: within one batch each request keeps the accepted
prefix and rolls back the rest. Verification is not. The engine's default,
which every measurement here uses, verifies the full number of tokens for every
request whatever its drafter proposed, so the verification pass has the same
shape at every step. DSpark also offers confidence-scheduled verification, a
non-default mode that verifies fewer tokens for some requests, chosen with a cost table
measured on the GPU~\cite{deepspec2026}. On the H100 the table prices an extra
verified token at a small fraction of a percent of a step, so the mode
verifies the full number of tokens at every batch. Acceptance is then unchanged at every batch
and the saving at serving batch falls by a few points, so all results use the
default mode.
Each drafter verifies the number of tokens it was released with, so that number differs across schemes
(Table~\ref{tab:drafters}). A block drafter verifies its trained block, and the
engine does not accept a shorter one. Where a block drafter and a head share a
model, the block drafter therefore also verifies more tokens per request,
and the comparisons in Section~\ref{sec:drafter} include that difference.
\begin{table}[t]
\centering
\caption{Drafters per model, given as the number of tokens each verifies per request and step and marked by who trained it (c with the model, v by the model's vendor afterwards, as Google's assistant for the gemma models, t by a third party), with the GPU each model's batch sweep runs on, and asterisks marking Mixture-of-Experts models.}
\label{tab:drafters}
\footnotesize
\setlength{\tabcolsep}{3pt}
\input{tab_drafters}
\end{table}
DeepSpec's DSpark drafter for gemma-4-12B-it has no implementation in the serving engine, because its layers are Gemma-4 layers rather than the Qwen3 layers the
engine's DSpark drafter builds. We wrote that implementation from the DeepSpec reference code; at a batch of 1 it reproduces the baseline's greedy output
token for token.

The counter updates more slowly than a single draft or verify phase lasts, so
we never divide energy between phases by time; every energy number is a total
over a whole generation, and the draft's share comes from removing the draft
(Section~\ref{sec:verify}). Production engines capture the decode step in CUDA
graphs wherever shapes allow, so deployments run with capture enabled, not in
eager mode. Capture
and speculation both exist to amortise kernel launches. A non-speculative decode
step is the most launch-bound thing the engine does, while a speculative step
already carries many tokens of work per launch, so capture helps the baseline more than it helps the speculative run and leaves speculation with less to
recover. Measuring with capture disabled
therefore credits speculation with an overhead it never removed.

We measured that credit by re-running the baseline and the speculative run with capture enabled. The credit is not a constant and cannot be transferred
between models, because it tracks how launch-bound the baseline is. The credit splits by architecture: on every dense model it is under 10 percentage points, and on every Mixture-of-Experts model it is 19 to 39 points at single stream (Table~\ref{tab:graphon}). Mixture-of-Experts
decode issues many small per-expert kernels and is the most launch-bound
baseline in the study. With
the launches removed, a Mixture-of-Experts step costs what its experts cost to
read, and every token in the step routes to the experts chosen for it, whether it
belongs to another request or is another drafted token. Verifying a draft
then reads more expert weights even for a single request.
Because the credit differs between schemes, capture can also change which scheme saves the most energy: at single stream on Qwen3.6-35B-A3B and gemma-4-26B-A4B-it, DFlash saves the most with capture disabled and the draft head saves the most with capture enabled, and n-gram drafting, which saves energy on every model with capture
disabled, costs energy on every Mixture-of-Experts model it runs on with
capture enabled.
Independent work reaches the same conclusion about Mixture-of-Experts decode
being launch-bound rather than arithmetic-bound, for an unrelated set of
optimisations~\cite{launchbound2026}.

We therefore report every saving, batch sweep, latency and energy split with
capture enabled (Table~\ref{tab:headline}), and use capture disabled only for
the hardware-counter roofline (Section~\ref{sec:roofline}). Graph replay
launches the same kernels, so the arithmetic and memory traffic each kernel
performs does not depend on capture.

\begin{table*}[t]
\centering
\caption{Largest energy saving among the schemes (\%) per model and batch with CUDA graph capture enabled, on the GPU given in Table~\ref{tab:drafters}, as the median over repeated runs $\pm$ half their range, the scheme that achieves it (DF DFlash, DS DSpark, MTP an MTP head or gemma's assistant), and a dash where no speculative run held the batch.}
\label{tab:headline}
\footnotesize
\setlength{\tabcolsep}{4.3pt}
\input{tab_headline}
\end{table*}

\section{What Speculation Saves, and Where the Saving Ends}
\label{sec:savings}
At single stream every model has a scheme that saves energy, and the largest saving per model, taken over its schemes, has a median of 54\%
of decode energy (Table~\ref{tab:headline}). The one
scheme that costs energy at single stream is n-gram drafting on the
Mixture-of-Experts models. N-gram drafting accepts too few tokens to pay for
the expert weights each extra verified token reads
(Section~\ref{sec:measurement}), while on the dense models the same drafter
saves energy.

\begin{figure}[t]
\centering
\includegraphics[width=\columnwidth]{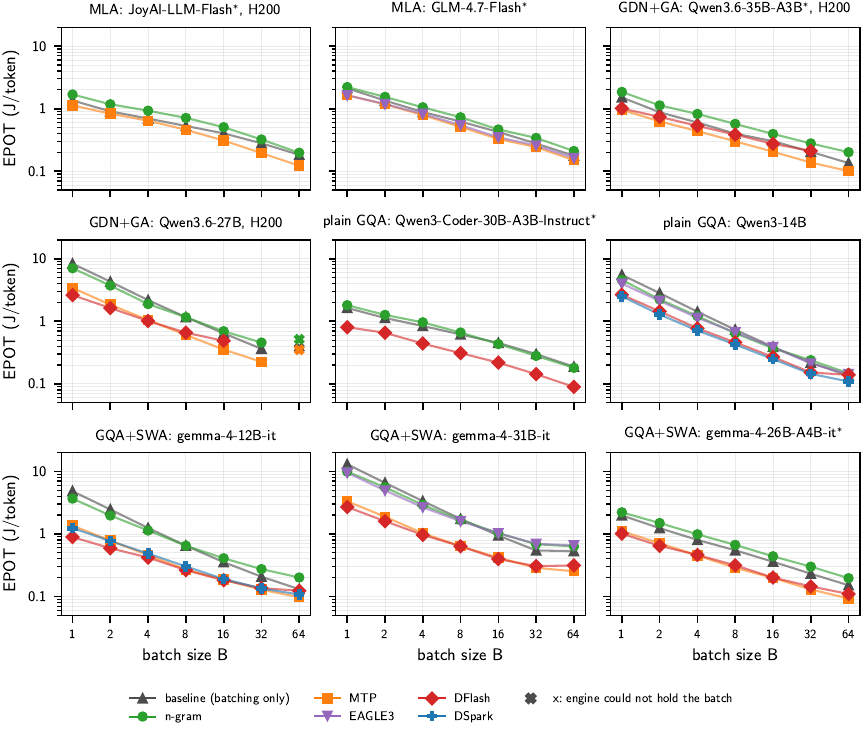}
\caption{Energy per output token against batch with CUDA graph capture enabled for 9 of the 11 models, grouped by attention family and labelled where the panel ran on the H200, with the non-speculative baseline in grey, asterisks marking Mixture-of-Experts models, crosses at batches the engine ran below the requested concurrency, and lines that stop short ending where the engine could not start.}
\label{fig:frontier}
\end{figure}

The saving then decays as batch grows (Figure~\ref{fig:frontier}), because
batching has already amortised the weight read that the drafter would divide.
This decay holds for the short contexts we serve; at long contexts the
key-value cache keeps large-batch decode memory-bound, and the speedup from
speculation can grow with batch~\cite{MagicDec}.

\begin{table*}[t]
\centering
\caption{Energy per output token of the non-speculative baseline and of the scheme with the largest saving with capture enabled, with CUDA graph capture off and on, and the saving each pair yields as the median of per-seed savings over three seeds with every speculative run paired to the baseline of the same seed and capture state, from a separate campaign run on the H100 except for JoyAI-LLM-Flash on the H200, so the Qwen3.6 rows differ from Table~\ref{tab:headline}.}
\label{tab:graphon}
\small
\input{tab_graphon}
\end{table*}

At a serving batch of $64$, the scheme with the largest saving on each model
still saves energy on every model we could run there, by 15 to 52\%
(Table~\ref{tab:headline}). Not every scheme saves energy at that batch: the n-gram drafter costs energy on every model we report it on except
Qwen3-Coder-30B-A3B-Instruct. On the dense Qwen3 models, the models closest to the two dense grouped-query
models of Fernandez et al.~\cite{fernandez2025energy}, DFlash-16 costs energy on
Qwen3-4B and Qwen3-8B, and EAGLE3 and DFlash-7 cost energy on Qwen3-14B, which
reproduces the sign change they report for two such models, while DSpark on all
three saves 18 to 24\%. The spread across models at fixed batch is wider than the spread
across schemes on most single models, so the model determines the saving
more than the drafting scheme does.

All batch claims stop at $64$, because beyond that our prompt pool could not
fill a batch with distinct prompts, and duplicate prompts share a cached
prefix, so such a batch would measure prefix reuse rather than batching.

\subsection{Does the Served Text Change the Answer?}
\label{sec:task}
A drafter predicts text, so the served content changes what it can recover. Structured, repetitive text is the favourable case: EAGLE reports its largest speedups on code and attributes them to fixed templates~\cite{li2024eagle}. On the Spec-Bench
cross-task benchmark, drafting by lookup in the prompt does best where the
output copies the input, as in summarisation, and almost every method gains
least on translation~\cite{xia2024specbench}. We ran the whole batch sweep on a code pool, a long-document pool and real
conversational turns, with every speculative run compared with a non-speculative baseline run on the
same pool at the same batch size, and with each pool built to avoid the
duplicate-prompt problem above.

Conversation is harder to predict than code for every drafter: every scheme
on every model accepts fewer tokens per verification on conversation, the
training-free n-gram drafter included. How much a drafter loses follows how it
was trained. The co-trained MTP heads lose little, and on GLM-4.7-Flash the largest saving, from its MTP head, is the same on both
pools. The post-hoc drafters lose more, and the
most specialised loses most: the DFlash drafter trained for
Qwen3-Coder-30B-A3B-Instruct accepts 2.0 to 3.0 tokens per verification on
conversation against 5.9 on code, and at a batch of $64$ it saves 51\% of
decode energy on code and costs energy on conversation. Across models, the largest saving per model at a batch of $64$ is a median of
11 percentage points
lower on conversation than on code. On long documents, where the answer copies
from the prompt, n-gram acceptance on the gemma models is higher than on
code, as Spec-Bench reports for lookup drafting.

The scheme that saves the most energy can change with the text, because a
drafter that accepts fewer tokens still pays for the same number of verified
tokens: the engine verifies the full number of tokens for every request,
whatever that request accepts (Section~\ref{sec:measurement}). On
gemma-4-12B-it and gemma-4-31B-it, the two schemes that save the most are the
DFlash drafter, which verifies 16 tokens per request and step, and the
assistant, which verifies 6 tokens. On code the DFlash drafter saves the most at a batch of 1
to 4 and the assistant at a batch of 32 and 64, the change with batch that
Section~\ref{sec:drafter} explains. On conversation the DFlash drafter accepts
3.3 to 3.4 tokens per verification, and the assistant accepts 3.3 
to 3.5 tokens: both drafters accept about as many tokens, the DFlash drafter
still verifies 16 tokens per request, and the assistant saves the most at every
batch. 

\begin{figure}[t]
\centering
\includegraphics[width=\columnwidth]{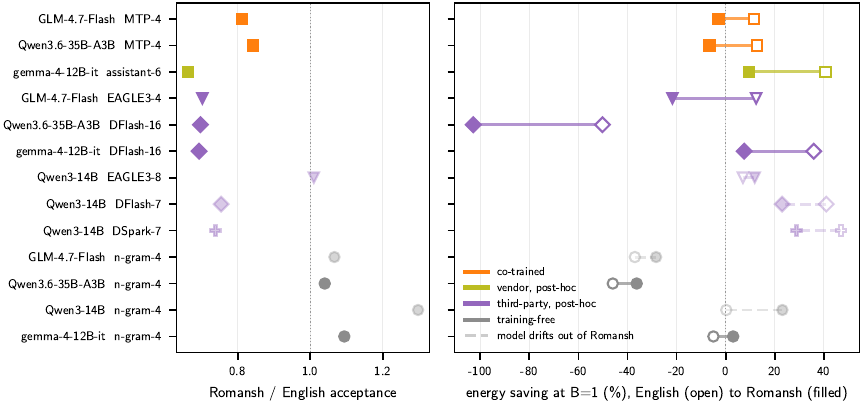}
\caption{Romansh-to-English acceptance ratio and energy saving at a batch of 1, one row per model and drafter labelled with the number of tokens it verifies per step, coloured by how the drafter was trained, and drawn light where the model does not answer consistently in Romansh.}
\label{fig:langtransfer}
\end{figure}

Spec-Bench attributes the weak translation speedups to thin multilingual data
in pretraining~\cite{xia2024specbench}, without separating how the drafter was
trained from how the model was trained. A language outside the drafters' training data separates how the drafter was trained from how the model was trained, and there are two common explanations: that block drafters
are fragile, and that what matters is when and on what data a drafter was
trained. We served matched news prompts in English and in Romansh, a low-resource
language. No model lists Romansh among its supported languages~\cite{qwen3,gemma4,card_glm47,card_joyai,card_qwen36},
and no post-hoc drafter's published training set contains it: those sets are English
instruction and chat data, and the z-lab DFlash recipe adds Spanish, French,
German, Italian and Japanese~\cite{ckpt_eagle3_glm,chen2026dflash,deepspec2026}.
Every model nevertheless writes Romansh, so some reached their pretraining,
and a co-trained head saw what its model saw. Each model was asked to
continue the article in the article's language, and only the responses that
stayed in Romansh, judged by the rate of Romansh function words that Italian
and the other neighbouring languages lack, are scored: 7 to 11 responses per
model and drafter at a batch of 1 (Figure~\ref{fig:langtransfer}).

On the 3 models where drafters of different origin serve the same model, the
outcome follows when each drafter was trained, not how it drafts or who trained it.
GLM-4.7-Flash and Qwen3.6-35B-A3B each ship a co-trained MTP head and are also
served by a post-hoc drafter, EAGLE3 on GLM-4.7-Flash and DFlash on
Qwen3.6-35B-A3B. On both, the head keeps 81 to 84\% of its English accepted length in
Romansh, and the post-hoc drafter 70\%. With capture enabled
these two Mixture-of-Experts models save little at single stream even in
English, so in Romansh every drafter on them costs energy, in particular post-hoc
drafters cost the most. On GLM-4.7-Flash the MTP head
and the EAGLE3 drafter both verify 4 tokens, and EAGLE3 drafts token by
token, so drafting a block at once is not the cause. On gemma-4-12B-it, the
assistant Google trained keeps no more of its English accepted length than
the third-party DFlash drafter, 66 to 69\%, and both lose most of their
saving: trained after its model, the assistant behaves as a post-hoc drafter
even though the model's vendor trained it. Qwen3-14B's DeepSpec drafters keep
most of their accepted length, but the model drifts out of Romansh into Italian on
23 to 27\% of its responses, which leaves it the smallest clean sample. A post-hoc drafter therefore needs to be re-measured on the text it will serve.

\section{How Much Batching Does It Replace?}
\label{sec:isobatch}
Every saving in the sections above compares a speculative run with a
non-speculative baseline run at the same batch size, so both runs serve the same number of
requests at once. That comparison shows what the drafter changes when the load
is fixed. A deployment can also lower energy per token without a drafter, by raising the batch size of the non-speculative baseline. This section therefore compares runs at
different batch sizes: for a speculative run at a given batch size, we find
the smallest batch size at which the baseline uses no more energy per token
than that speculative run. On each model, the speculative run is the scheme
with the lowest energy per token at that batch size, and the baseline batch
sizes are the ones we ran, 1 to 64 in powers of 2.

A speculative run serving one request uses as little energy per token as the
baseline serving 2 to 8 requests, depending on the model. Against a
speculative run serving 32 requests, the baseline needs at least 64 requests
on every model, and on 5 of the 11 models the baseline serving 64 requests, the
largest batch size we ran, still uses more energy per token than the
speculative run serving 32. The batch size cannot be
raised past what latency permits, and that limit is what motivates
speculation.

\begin{figure*}[t]
\centering
\includegraphics[width=\textwidth]{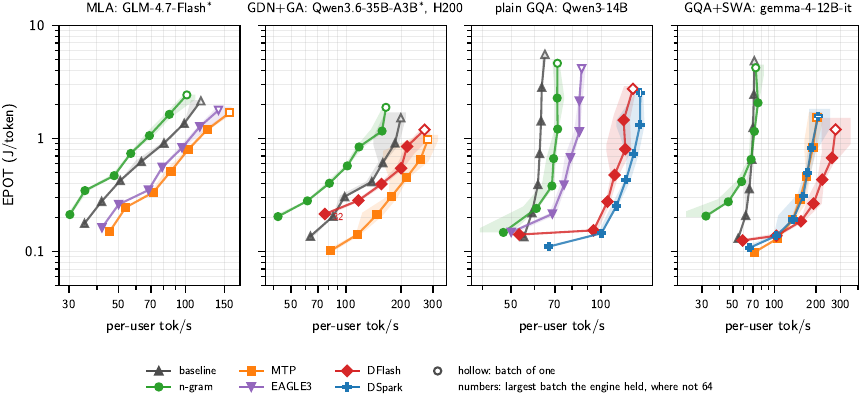}
\caption{Energy per token against per-user throughput with CUDA graph capture enabled, one panel per attention family, with each trace running from a batch of 1 (hollow marker) to a batch of 64 or to the batch given in its label, shaded over the range of its three seeds.}
\label{fig:speedenergy}
\end{figure*}

Putting speed and energy on the same axes shows why the substitution is
partial (Figure~\ref{fig:speedenergy}). As the batch of the baseline grows, its trace falls almost vertically: raising the batch size cuts energy per token
by more than an order of magnitude. On the dense models each user is served at
close to the same rate throughout; on the Mixture-of-Experts models the trace
also leans left, because each added request adds expert reads to every step and
slows every user. Speculation moves the trace to the right instead, raising
per-user speed at single stream by $1.3$ to $4.3\times$. On dense models the
two levers are close to perpendicular. Batching is the energy lever and costs
little latency; speculation is the latency lever, and the energy it also buys
is the part that decays with batch. For a service that is not
latency-constrained, batching is the cheaper way to reduce energy; for one that
is, speculation provides the per-user speed that batching cannot, and on a
Mixture-of-Experts model it also recovers speed that batching gives up.

Table~\ref{tab:tbt} gives the mean time between tokens behind
Figure~\ref{fig:speedenergy}. Under an objective of $25$\,ms per token, the
lowest-energy configuration that meets it within the batches we ran uses 18 to
69\% less energy per token with speculation than without on 10 of the 11
models, and gemma-4-31B-it meets it only with speculation. The time is a mean
over each generation; per-token percentiles need timestamps that our runs do
not record.

Lowering the GPU clock through dynamic voltage and frequency scaling (DVFS)
offers a third lever, one that keeps latency
where the step waits on memory. We locked the SM clock of an H200 at 1500,
1200 and 900\,MHz for Qwen3-8B on the code pool,  ran with three seeds per batch, and
compared each lock with the default power limit. Without speculation the
baseline step runs near the top clock while it waits on memory, so the
1500\,MHz lock costs 3 to 7\% of throughput and lowers energy per
token by 4 to 11\% at every batch. With DSpark-7 or DFlash-16 at a batch of 16
or more, the GPU already runs at between 1450 and 1700\,MHz under
the default limit, so the same lock removes little clock and saves no
energy. The 1200\,MHz lock lowers energy for every configuration but also lowers
throughput. The up to 32\% that clock locking recovers on the H200
in~\cite{ma2026powercap,sweetspots2026} comes from the lock that minimises
energy, on other models and without CUDA graph capture.

\begin{table*}[t]
\centering
\caption{Mean time between output tokens per user in milliseconds with CUDA graph capture enabled on the code pool, for the non-speculative baseline and for the speculative scheme with the shortest time at each batch, with a dash where the engine did not hold the batch and a dagger where prefill time is not subtracted.}
\label{tab:tbt}
\footnotesize
\input{tab_tbt}
\end{table*}

\section{Which Drafter, and Does It Change With Batch?}
\label{sec:drafter}
The drafter that saves the most energy at single stream is not the drafter
that saves the most at serving batch. At single
stream a block drafter leads on 7 of the 9 models that have a block drafter (Table~\ref{tab:headline}), and on all 9 models with capture
disabled, because block drafters accept the most tokens per verification. On every model that also has a draft
head, a co-trained MTP head on the Qwen3.6 models and the assistant on the
gemma models, the head saves the most energy at the largest batch the model runs.
On Qwen3.6-27B, gemma-4-12B-it and gemma-4-31B-it the head takes over between
a batch of 8 and a batch of 32, on gemma-4-31B-it with DFlash leading again at a
batch of 16; on Qwen3.6-35B-A3B it saves the most at every
batch, and on gemma-4-26B-A4B-it at every batch except 2.

The cost that grows with batch is the number of tokens verified. A block
drafter proposes all its tokens in one pass: the z-lab DFlash drafters propose 15 tokens, so the model verifies 16 tokens per request and step. The draft heads on these models, the MTP heads and the assistant, propose one
token per pass and stop after 3 to 5 tokens, so the model verifies 4 to 6
tokens per request and step. The model verifies every proposed token for every
request, accepted or not. Rejected tokens cost little at single stream, where
the step waits on memory, and add compute at serving batch
(Section~\ref{sec:background}), so a drafter that proposes more tokens per
step adds more of that compute. The draft itself does not grow with batch (Section~\ref{sec:verify}).
The head starts to save the most energy at the batch where the energy of verifying
16 tokens instead of 4 to 6 tokens exceeds the energy saved by the extra
tokens the block drafter gets accepted.

We modified the engine so that the n-gram drafter can serve the two
state-space hybrids, Qwen3.6-27B and Qwen3.6-35B-A3B. A hybrid layer carries a
recurrent state, and after each verification the engine must advance that
state to the last accepted token. The MTP and DFlash code paths of the engine
perform this step, but the n-gram path does not, so the state stops advancing
after the first verification and the output becomes incoherent within the
first few words. With the step added to the n-gram path,
n-gram text departs from the baseline text no earlier than MTP text does. On
the models without a recurrent state, n-gram text departs from the baseline
text no earlier than the texts of two baseline runs depart from each other.

On the hybrids, n-gram drafting then follows the trend it shows on the models
without a recurrent state (Figure~\ref{fig:frontier}). On the dense Qwen3.6-27B it saves energy up to a batch of
4, 12\% at a batch of 1, breaks even at a batch of 8 and costs energy from a
batch of 16. On the Mixture-of-Experts Qwen3.6-35B-A3B it costs energy at
every batch (Section~\ref{sec:savings}).

Among block drafters, the drafter that proposes fewer tokens per pass saves
more energy. Qwen3-4B and Qwen3-8B each have two DFlash drafters, trained by
different groups. The DeepSpec drafter proposes 6 tokens per pass, so the model verifies 7 tokens per request and step; the z-lab drafter proposes 15 tokens, so the model verifies 16 tokens. The drafter that proposes 6 tokens saves more
energy at every batch we measured, and its margin grows with batch. The model
verifies the 9 extra tokens for every request, and at large batch the
extra accepted tokens no longer cover the compute those 9 tokens add.

Memory can also cap the batch, independent of energy. On a state-space
hybrid, rolling back a rejected draft is not a matter of discarding cache
entries: the recurrent state has already absorbed the rejected tokens, so the
engine keeps a copy of that state after every drafted token of every
request, and restores the one that matches the accepted prefix. The buffer
grows with the number of drafted tokens times the batch. On Qwen3.6-27B on the H100, DFlash-16
fails to start at a batch of 16, which the baseline runs, rather than
running slowly; at that batch the rollback buffer alone is about 10 times the
size of the drafter's weights. The H200's larger memory moves the ceiling but
does not remove it: DFlash-16 still fails to start at the largest batches on
both hybrids, while the co-trained MTP head, which verifies 4 tokens, holds
Qwen3.6-35B-A3B at a batch of $64$. A deployment on a recurrent architecture should budget memory for the number of drafted tokens, not only for the
weights of the drafter.

\subsection{How Many Tokens Should a Draft Hold?}
\label{sec:kstar}
The same argument sets the number of tokens to draft. Drafting
more tokens per verification helps when they are accepted, because more
tokens come out of one weight read, and wastes draft compute when they are
not. Acceptance falls off as the draft runs further ahead of the model, so
there is an interior optimum.

Batch moves that optimum downward, because the verify pass already carries
many tokens while each drafted token is still verified for every request.

We swept the number of tokens the n-gram drafter proposes from 1 to 32 tokens at 4 batch sizes on gemma-4-31B-it, Qwen3-Coder-30B-A3B-Instruct and
gemma-4-26B-A4B-it. On gemma-4-31B-it, the dense model of the three models, drafting
more tokens saves more energy at single stream, but from a batch of 4 requests 4 drafted tokens save the most energy, and at serving batch every number of
drafted tokens costs energy, 32 tokens the most. On the two Mixture-of-Experts models every number of drafted tokens costs energy already at single stream,
and the loss grows with the number of tokens, because each verified token reads
the experts chosen for that token (Section~\ref{sec:measurement}). On all three models the least costly choice at a batch of $64$ is 4 tokens.
For the n-gram drafter, and for the two DFlash drafters of
Section~\ref{sec:drafter}, over-drafting, not under-drafting, is the failure mode, and a number of drafted
tokens tuned at single stream on a dense model should be cut when the batch is
raised.

\section{Which Metric Predicts the Energy Ranking?}
\label{sec:predictor}
\begin{figure*}[t]
\centering
\includegraphics[width=\textwidth]{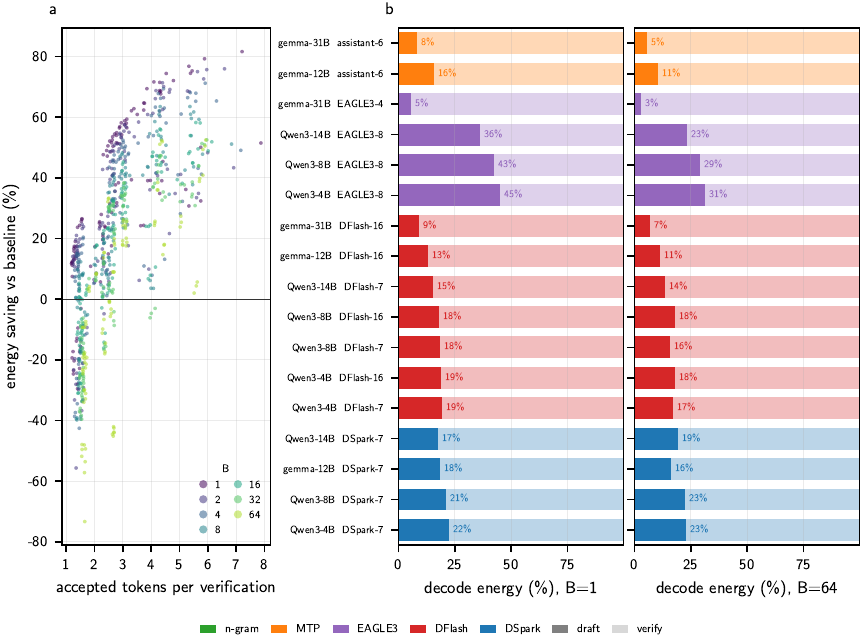}
\caption{Energy saving against accepted tokens per verification for every model, scheme and batch that has a baseline run at the same batch size, coloured by batch (a), and the share of decode energy spent drafting and verifying at $B{=}1$ and $B{=}64$ with capture enabled for every dense model and scheme, measured by removing the draft (b).}
\label{fig:tune}
\end{figure*}

Acceptance, the number of drafted tokens the model keeps per verification, is
the obvious metric. Drafter
training optimises acceptance, and every drafter release reports it.
Acceptance alone does not tell which drafter uses less energy, because that
drafter changes with batch while acceptance does not. For any one drafter,
acceptance changes little from a batch of 1 to a batch of $64$. Block
drafters accept more tokens than draft heads in every matched comparison we
measured, at every batch. Yet on all 5 models that have both, the head uses less energy at the largest
batch, because the block drafter also verifies more tokens
(Section~\ref{sec:drafter}); acceptance counts only the tokens kept.

Once the batch is fixed, acceptance does predict the saving
(Figure~\ref{fig:tune}a). At every batch the saving rises with the number of
accepted tokens per verification, the quantity that divides the weight read.
The acceptance needed to break even rises with batch, because batching has
already amortised the weight read that a drafter would otherwise divide. At a
fixed batch, the scheme that accepts more also uses less energy in $287$ of
$332$ pairs of schemes on the same model; the pairs share models and baselines,
so they are not independent tests. In all but one of the $45$ pairs
where it does not, the scheme that accepts more also verifies more tokens per
request, most often a block drafter against a draft head. The remaining pair
is EAGLE3 on Qwen3-8B at single stream, which drafts 7 tokens in 7 sequential passes.

\section{Where the Energy Goes}
\label{sec:verify}
Verification takes most of decode energy on every dense model and scheme, at
single stream and at a batch of $64$ (Figure~\ref{fig:tune}b). We measure the
split with capture enabled by removing the draft. The speculative run is
repeated with the drafter's passes skipped after the first step and the first
step's draft verified at every following step, so verification keeps its number of tokens and
shape. An engine setting fixes how many drafted tokens each step accepts in both
runs, so
they run the same steps over the same contexts, and the difference in their
decode energy is the energy of the draft. On a dense model the tokens verified in place of
a draft do not matter: on Qwen3-8B, verifying random or repeated tokens instead
changes the energy by under 3\%.

No drafter takes more than 45\% of decode energy, and a block drafter takes at
most 23\%, so the cost of running the drafter is the smaller lever: halving the
cost of a block drafter would lower decode energy by at most 12\%. The share
follows how many sequential passes a drafter runs and how large each pass is
next to a pass of the model. EAGLE3 on Qwen3 runs 7 passes per draft and
takes the largest share, while a block drafter runs one pass per block. On
gemma-4-31B-it, the largest dense model, the passes of every drafter are small next
to those of the model, and every share is small. The share stays close to constant or falls
from single stream to a batch of $64$, because verification grows with the
batch at least as fast as the draft does.

Phase markers emitted by the host, the usual way to split the two phases,
need capture disabled and do not give this split on the gemma models
(Section~\ref{sec:threats}). With capture disabled, the markers credit the
gemma-4-12B-it assistant at a batch of $64$ with a larger share of decode
energy than removing the draft measures. 

On the Mixture-of-Experts models the subtraction does not hold, because each
verified token is routed to the experts chosen for that token. On GLM-4.7-Flash, verifying random
or repeated tokens in place of the reused draft changes the energy of the
verify-only run by up to 52\%. On these models the cost of
verification depends on which tokens are drafted and not only on how many, so
we do not report their split.

\section{Does Speculation Reach the Compute-Bound Regime?}
\label{sec:roofline}
The standard explanation of speculation is a roofline shift: the step is said
to move toward the regime where the machine's arithmetic units
are the limit rather than its memory. If that were happening, energy per token
would fall for a structural reason, and the effect would grow with the number of
tokens verified.

We measured the operating point with hardware counters rather than inferring
it, taking executed arithmetic intensity as measured arithmetic, including the
padded work of tensor-core tiles, over measured memory traffic for a whole
decode step, on an H100 and on an H200.  The counters capture the first decode steps after the prompt, so the
context is set by the prompt length: we use prompts of 1024 tokens and of
about 260 tokens, the average context over a 512-token generation from our
code prompts.

\begin{figure}[t]
\centering
\includegraphics[width=\columnwidth]{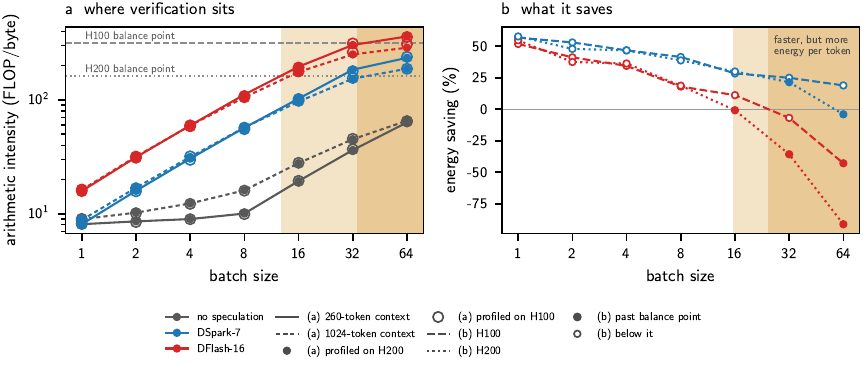}
\caption{Arithmetic intensity of a decode step on Qwen3-4B against batch, with the balance point of each GPU as a horizontal line (a), and the energy saving of the same drafters with capture enabled on each GPU, filled where the step at that batch is past the balance point of that GPU (b), shaded in (a) from the batch at which the DFlash-16 step crosses each balance point and in (b) from the batch at which its saving turns negative on each GPU.}
\label{fig:roofline}
\end{figure}

The number of verified tokens alone moves the operating point little: at single
stream every scheme on every model sits within a few per cent of where its
baseline sits. Combined with batch, it moves the operating point a long way. Arithmetic intensity is
a property of the step and not of the GPU, and the same step measures close
to the same on both (Figure~\ref{fig:roofline}a). With 260-token contexts, a
DFlash-16 step on Qwen3-4B reaches the balance point of the H100 at a batch of $32$
and that of the H200 at $16$. DSpark-7, which verifies fewer than half as many
tokens, stays below the H100's balance point at every batch and passes the
balance point of the H200 from a batch of $32$. The baseline stays far below both balance points
at every batch. The number of verified tokens sets the crossing, not the drafter: the n-gram
drafter verifying 16 tokens passes the balance point of the H200 at a batch
of $16$, as DFlash-16 does, while verifying 4 tokens, as in our panels, it
stays below it up to a batch of $64$. Longer contexts lower the intensity at a given batch, because
each request's context has to be read as well, which adds more memory traffic
than arithmetic.

Below the balance point, the n-gram drafter and the assistant on
gemma-4-12B-it run below the memory roof from a batch of 1, because attention
over their drafted tokens runs in slow kernels, while DSpark runs close to it.

Reaching the compute-bound regime is not a gain in energy; it marks where the
saving ends (Figure~\ref{fig:roofline}b). Past the balance point, every
verified token costs arithmetic time that the weight read no longer hides,
and most verified tokens are rejected: DFlash-16 accepts under 3 of the 16 tokens it verifies per request. On both GPUs and for both drafters, the
saving changes sign within one doubling of the batch at which the step reaches
that GPU's balance point. DFlash-16 stops saving between batches of $16$ and
$32$ on the H100 and at $16$ on the H200. DSpark-7 still saves energy at a
batch of $64$ on the H100, where its step stays below the balance point, and
costs energy on the H200, where its step passes the balance point. A GPU with more memory bandwidth per unit of arithmetic reaches the
compute-bound regime at a smaller batch and loses the saving there, so a number of
drafted tokens tuned on one GPU does not carry over to a GPU with a different balance
point. The H200 also changes the energy of the baseline itself, through its
memory and power behaviour, so the balance point explains where the saving
changes sign, not its size.

\section{Measurement Pitfalls}
\label{sec:threats}
Speculative decoding energy is easy to mismeasure, and the failures share one
form: the run succeeds, the result file is well formed, the number is
plausible, and nothing in it shows that the configuration differed from the
one requested.

Besides the capture bias of Section~\ref{sec:measurement}, the following
conditions produce the same failure. The serving engine resolves concurrency separately for the baseline and the speculative run, so the batch that ran can differ from the batch
requested, and the result file stores the requested batch and the batch that ran in
different fields. A batch larger than the prompt pool repeats prompts that share a cache prefix
(Section~\ref{sec:savings}). A prompt sent to an instruction-tuned model without its chat template
changes the text being predicted and moves the reported saving in either
direction: without the template, n-gram drafting on the gemma models reports
savings up to 117 percentage points higher, and the MTP head on
JoyAI-LLM-Flash up to 12 points lower. A profiler window that opens
while requests are still being admitted includes prefill, which the decode
analysis excludes. Phase markers emitted by the host record when the host enters a
phase, not when the GPU runs it, so they can book GPU work of verification to drafting
(Section~\ref{sec:verify}). A result writer that matches each appended row to the
file's existing header discards, without warning, any column the header
lacks, including the column that records the condition a row was measured
under. Result rows that do not name the GPU can mix H100 and H200 runs. The same requested batch can then run in full on one GPU and below the
requested concurrency on the other GPU: the MTP head on
Qwen3.6-27B at a batch of $16$ holds 12 requests on the H100 and saves a
fraction of what it saves on the H200, which holds all 16 requests.

Holding the generation length fixed keeps the baseline and the speculative
run on the same token count, but a run that continues past the end of the model's answer starts a new
turn and repeats text. That repetition does not cancel within
the pair. The baseline pays the same per token whatever it generates, while a
drafter predicts repeated text cheaply, so acceptance rises once the answer
has ended.
With the end-of-sequence token honoured, every model answers our code prompts
in a median of more than 700 tokens, so the 512 tokens of every run stay inside
the answer.

A speculative run can produce broken text while its result row looks normal,
as n-gram does on the state-space hybrids without the engine fix
(Section~\ref{sec:drafter}). Taken at face value, those runs would show n-gram
matching or beating the co-trained head on Qwen3.6-27B up to a batch of 8.

Each condition above shows only outside the reported metric: in the generated
text, the count of distinct outputs, the engine log or the profiled kernel
count. The minimum audit confirms the workload, the concurrency and the
measurement window, and reports the capture mode and the GPU with every number.

Our measurements have limits: latency is the mean time between tokens, the
counter covers the GPU board only, prompts are short, batches stop at 64, heads
and block drafters differ in the number of tokens they verify, only the n-gram
drafter has its number of drafted tokens swept, and the roofline covers Qwen3-4B.

\section{Related Work}
\label{sec:related}
\emph{Drafting schemes.} Speculative decoding was introduced with a separate,
smaller network as the drafter~\cite{leviathan2023fast}. The drafters in wide use
remove that separate network: Medusa~\cite{cai2024medusa} attaches parallel heads, EAGLE and
EAGLE-3 draft over the hidden features of the model~\cite{li2024eagle,li2025eagle3}, and several recent checkpoints ship
multi-token-prediction heads trained jointly with the
model~\cite{deepseekv3}. Block drafters emit a whole block in one
pass, and the checkpoints we use are of that
family~\cite{deepspec2026,ckpt_dflash_zlab,ckpt_dspark_qwen3_14b}.  All of
this work is evaluated in accepted length and wall-clock speedup. We ask what
the same drafters cost in joules, which orders them differently once the batch
is non-trivial.  Serving systems tune the number of drafted tokens in
closed loop against throughput~\cite{liu2024turbospec}.

\emph{Energy measurement.} Roofline analysis explains why decode is
memory-bound~\cite{williams2009roofline}, and phase-attributed energy
benchmarks exist for the prefill and decode
split~\cite{tokenpowerbench2026,joules2026}, and for power-capped
operation~\cite{ma2026powercap}. Two studies are closer to ours: decode energy has been
measured across attention mechanisms with hardware counters, without
speculation~\cite{energyattn2026}, and the energy savings of speculative
strategies have been surveyed on older dense
models~\cite{specenergy2026}. None separates draft from verify energy, and that
split determines whether a drafter pays for itself.

\emph{Positioning.} Closest to this work, Fernandez et
al.~\cite{fernandez2025energy} measure speculative decoding's energy on two dense grouped-query models with a single drafter at fixed draft length. We
measure across 4 attention families and 5 schemes, vary the number of
drafted tokens, and decompose the energy.  The same comparison of schemes has been measured in time on a
production engine~\cite{spec-illusion} and in throughput by a dedicated
benchmark~\cite{speedbench2026}; neither reports joules. An analytic roofline
treatment of architecture-dependent speculation predicts several of the
effects we measure~\cite{finn2026speculation}, and SpecLA~\cite{specla2026}
builds verification kernels for linear-attention states; we measure the
energy of speculation on such models. That Mixture-of-Experts models gain
more from speculation than dense ones is an established
result~\cite{huang2025moesd,coheremoespec2026}; we show that the launch-bound decode behind that gain also makes their
speculative energy the most sensitive to graph capture.

\emph{Language and task.} With a separate small drafter, speculative decoding is
less effective in low-resource languages, and distilling that drafter on the
target language generalises poorly~\cite{paudel2026multilingual,sandler2025disparate}.
Pretraining a drafter on broad data helps most on out-of-domain and
multilingual text~\cite{bie2026osprey}. None of these compares, on one model, the different
drafters that production engines ship for it, and none measures energy.

\section{Conclusion}
\label{sec:conclusion}
Speculation shortens the time between output tokens, which batching does not,
and under a tight enough objective no batch size meets it without speculation,
so a service in that regime chooses speculation to meet its objective rather
than on energy grounds.

Speculation is most valuable for energy where a latency target rules out a
larger batch size, and least valuable where the batch size can still grow,
because there the drafter competes with batching, a lever that needs no extra
model and costs only cache memory and per-user latency. Most services value
time between output tokens above energy per token, and they should run
speculation and tune the drafter and the number of drafted tokens to the batch. Where energy per token is the
binding constraint, for instance under a facility power limit, a service
should raise the batch size first and then check whether a drafter still
reduces energy.

Three findings carry beyond our models. A draft long enough to push
verification past the GPU's balance point loses its saving there, and that
point depends on the GPU, so the same drafter at the same batch can save
energy on one GPU and cost energy on another with more memory bandwidth. The
drafter that saves the most energy at single stream is often not the drafter
that saves the most at serving batch, and acceptance, which changes little with
batch, does not show the change. Each condition in
Section~\ref{sec:threats} changed a reported number in this study, and none
was visible in the energy number.

\bibliographystyle{ACM-Reference-Format}
\bibliography{refs}

\end{document}

%% file: tab_drafters.tex
% Checkpoint facts, not measurements: one row per model, one column per
% scheme, each cell the positions verified per request and step and who
% trained the drafter. Source of truth: the arm definitions in
% runners/slurm_len512.sbatch (panels q4 q8 q14 q4e3 q8e3 coder g12 g12ds g26
% g31 g31e3 glm joyai q27 q35). Update both together.
\begin{tabular}{@{}lllccccc@{}}
\toprule
family & model & GPU & n-gram & MTP head & EAGLE3 & DFlash & DSpark \\
\midrule
GQA & \code{Qwen3-4B} & H100 & 4 & -- & 8\,\textsuperscript{t} & 16\,\textsuperscript{t}, 7\,\textsuperscript{t} & 7\,\textsuperscript{t} \\
    & \code{Qwen3-8B} & H100 & 4 & -- & 8\,\textsuperscript{t} & 16\,\textsuperscript{t}, 7\,\textsuperscript{t} & 7\,\textsuperscript{t} \\
    & \code{Qwen3-14B} & H100 & 4 & -- & 8\,\textsuperscript{t} & 7\,\textsuperscript{t} & 7\,\textsuperscript{t} \\
    & \code{Qwen3-Coder-30B-A3B-Instruct}$^{*}$ & H100 & 4 & -- & -- & 16\,\textsuperscript{t} & -- \\
GQA+SWA & \code{gemma-4-12B-it} & H100 & 4 & 6\,\textsuperscript{v} & -- & 16\,\textsuperscript{t} & 7\,\textsuperscript{t} \\
    & \code{gemma-4-26B-A4B-it}$^{*}$ & H100 & 4 & 6\,\textsuperscript{v} & -- & 16\,\textsuperscript{t} & -- \\
    & \code{gemma-4-31B-it} & H100 & 4 & 6\,\textsuperscript{v} & 4\,\textsuperscript{t} & 16\,\textsuperscript{t} & -- \\
MLA & \code{GLM-4.7-Flash}$^{*}$ & H100 & 4 & 4\,\textsuperscript{c} & 4\,\textsuperscript{t} & -- & -- \\
    & \code{JoyAI-LLM-Flash}$^{*}$ & H200 & 4 & 4\,\textsuperscript{c} & -- & -- & -- \\
GDN+GA & \code{Qwen3.6-27B} & H200 & 4 & 4\,\textsuperscript{c} & -- & 16\,\textsuperscript{t} & -- \\
    & \code{Qwen3.6-35B-A3B}$^{*}$ & H200 & 4 & 4\,\textsuperscript{c} & -- & 16\,\textsuperscript{t} & -- \\
\bottomrule
\end{tabular}

%% file: tab_headline.tex
% GENERATED by analysis/collect_results.py -- do not edit by hand.
\begin{tabular}{@{}llrrrrrrr@{}}
\toprule
family & model & $B{=}1$ & $B{=}2$ & $B{=}4$ & $B{=}8$ & $B{=}16$ & $B{=}32$ & $B{=}64$ \\
\midrule
GDN$+$GA & \code{Qwen3.6-27B} & +65{\tiny\,$\pm$5} {\scriptsize DF} & +61{\tiny\,$\pm$2} {\scriptsize DF} & +54{\tiny\,$\pm$1} {\scriptsize DF} & +48{\tiny\,$\pm$1} {\scriptsize MTP} & +45{\tiny\,$\pm$1} {\scriptsize MTP} & +38{\tiny\,$\pm$1} {\scriptsize MTP} & -- \\
 & \code{Qwen3.6-35B-A3B} & +38{\tiny\,$\pm$4} {\scriptsize MTP} & +31{\tiny\,$\pm$8} {\scriptsize MTP} & +26{\tiny\,$\pm$2} {\scriptsize MTP} & +25{\tiny\,$\pm$2} {\scriptsize MTP} & +31{\tiny\,$\pm$3} {\scriptsize MTP} & +31{\tiny\,$\pm$1} {\scriptsize MTP} & +25{\tiny\,$\pm$1} {\scriptsize MTP} \\
GQA & \code{Qwen3-14B} & +54{\tiny\,$\pm$1} {\scriptsize DS} & +53{\tiny\,$\pm$3} {\scriptsize DS} & +49{\tiny\,$\pm$2} {\scriptsize DS} & +41{\tiny\,$\pm$1} {\scriptsize DS} & +35{\tiny\,$\pm$1} {\scriptsize DS} & +34{\tiny\,$\pm$0} {\scriptsize DS} & +18{\tiny\,$\pm$1} {\scriptsize DS} \\
 & \code{Qwen3-4B} & +56{\tiny\,$\pm$5} {\scriptsize DS} & +53{\tiny\,$\pm$4} {\scriptsize DS} & +48{\tiny\,$\pm$1} {\scriptsize DS} & +41{\tiny\,$\pm$1} {\scriptsize DS} & +29{\tiny\,$\pm$2} {\scriptsize DS} & +25{\tiny\,$\pm$1} {\scriptsize DS} & +19{\tiny\,$\pm$1} {\scriptsize DS} \\
 & \code{Qwen3-8B} & +62{\tiny\,$\pm$3} {\scriptsize DS} & +55{\tiny\,$\pm$3} {\scriptsize DS} & +50{\tiny\,$\pm$2} {\scriptsize DS} & +41{\tiny\,$\pm$2} {\scriptsize DS} & +37{\tiny\,$\pm$1} {\scriptsize DS} & +34{\tiny\,$\pm$0} {\scriptsize DS} & +24{\tiny\,$\pm$1} {\scriptsize DS} \\
 & \code{Qwen3-Coder-30B-A3B-Instruct} & +32{\tiny\,$\pm$10} {\scriptsize DF} & +35{\tiny\,$\pm$5} {\scriptsize DF} & +41{\tiny\,$\pm$10} {\scriptsize DF} & +48{\tiny\,$\pm$7} {\scriptsize DF} & +49{\tiny\,$\pm$3} {\scriptsize DF} & +51{\tiny\,$\pm$1} {\scriptsize DF} & +51{\tiny\,$\pm$2} {\scriptsize DF} \\
GQA$+$SWA & \code{gemma-4-12B-it} & +75{\tiny\,$\pm$7} {\scriptsize DF} & +73{\tiny\,$\pm$4} {\scriptsize DF} & +65{\tiny\,$\pm$1} {\scriptsize DF} & +59{\tiny\,$\pm$3} {\scriptsize DF} & +49{\tiny\,$\pm$1} {\scriptsize DF} & +37{\tiny\,$\pm$1} {\scriptsize MTP} & +25{\tiny\,$\pm$1} {\scriptsize MTP} \\
 & \code{gemma-4-26B-A4B-it} & +39{\tiny\,$\pm$3} {\scriptsize MTP} & +41{\tiny\,$\pm$7} {\scriptsize DF} & +42{\tiny\,$\pm$1} {\scriptsize MTP} & +47{\tiny\,$\pm$2} {\scriptsize MTP} & +45{\tiny\,$\pm$1} {\scriptsize MTP} & +42{\tiny\,$\pm$2} {\scriptsize MTP} & +36{\tiny\,$\pm$1} {\scriptsize MTP} \\
 & \code{gemma-4-31B-it} & +77{\tiny\,$\pm$4} {\scriptsize DF} & +75{\tiny\,$\pm$3} {\scriptsize DF} & +68{\tiny\,$\pm$2} {\scriptsize DF} & +62{\tiny\,$\pm$1} {\scriptsize MTP} & +58{\tiny\,$\pm$2} {\scriptsize DF} & +47{\tiny\,$\pm$0} {\scriptsize MTP} & +52{\tiny\,$\pm$1} {\scriptsize MTP} \\
MLA & \code{GLM-4.7-Flash} & +22{\tiny\,$\pm$1} {\scriptsize MTP} & +12{\tiny\,$\pm$1} {\scriptsize MTP} & +12{\tiny\,$\pm$0} {\scriptsize MTP} & +18{\tiny\,$\pm$2} {\scriptsize MTP} & +22{\tiny\,$\pm$1} {\scriptsize MTP} & +12{\tiny\,$\pm$0} {\scriptsize MTP} & +15{\tiny\,$\pm$1} {\scriptsize MTP} \\
 & \code{JoyAI-LLM-Flash} & +15{\tiny\,$\pm$6} {\scriptsize MTP} & +6{\tiny\,$\pm$7} {\scriptsize MTP} & +5{\tiny\,$\pm$3} {\scriptsize MTP} & +14{\tiny\,$\pm$1} {\scriptsize MTP} & +23{\tiny\,$\pm$0} {\scriptsize MTP} & +30{\tiny\,$\pm$0} {\scriptsize MTP} & +33{\tiny\,$\pm$1} {\scriptsize MTP} \\
\bottomrule
\end{tabular}

%% file: tab_graphon.tex
% GENERATED by analysis/graphon_table.py -- do not edit by hand.
\begin{tabular}{@{}llrrrrrr@{}}
\toprule
 & & \multicolumn{2}{c}{baseline (J/token)} & \multicolumn{2}{c}{speculative (J/token)} & \multicolumn{2}{c}{saving (\%)} \\
\cmidrule(lr){3-4}\cmidrule(lr){5-6}\cmidrule(lr){7-8}
model & scheme & off & on & off & on & off & on \\
\midrule
\multicolumn{8}{@{}l}{$B{=}1$} \\
\code{gemma-4-31B-it} & DFlash & 12.80 & 12.65 & 2.96 & 2.91 & $+77$ & $\mathbf{+77}$ \\
\code{gemma-4-12B-it} & DFlash & 6.33 & 4.92 & 1.50 & 1.23 & $+76$ & $\mathbf{+75}$ \\
\code{Qwen3.6-27B} & DFlash & 11.60 & 10.51 & 3.43 & 3.04 & $+71$ & $\mathbf{+71}$ \\
\code{Qwen3-8B} & DSpark & 3.79 & 3.10 & 1.60 & 1.23 & $+57$ & $\mathbf{+60}$ \\
\code{Qwen3-4B} & DSpark & 2.88 & 1.77 & 1.23 & 0.757 & $+57$ & $\mathbf{+57}$ \\
\code{Qwen3-14B} & DSpark & 5.73 & 5.64 & 2.48 & 2.61 & $+57$ & $\mathbf{+54}$ \\
\code{gemma-4-26B-A4B-it} & MTP & 5.10 & 1.95 & 2.15 & 1.20 & $+58$ & $\mathbf{+39}$ \\
\code{Qwen3-Coder-30B-A3B-Instruct} & DFlash & 5.32 & 1.60 & 1.51 & 1.12 & $+71$ & $\mathbf{+32}$ \\
\code{Qwen3.6-35B-A3B} & MTP & 6.13 & 1.55 & 2.63 & 1.12 & $+56$ & $\mathbf{+30}$ \\
\code{GLM-4.7-Flash} & MTP & 8.02 & 2.12 & 4.48 & 1.73 & $+44$ & $\mathbf{+19}$ \\
\code{JoyAI-LLM-Flash} & MTP & 6.20 & 1.37 & 3.29 & 1.13 & $+47$ & $\mathbf{+18}$ \\
\midrule
\multicolumn{8}{@{}l}{$B{=}64$} \\
\code{gemma-4-31B-it} & MTP & 0.531 & 0.521 & 0.255 & 0.254 & $+52$ & $\mathbf{+51}$ \\
\code{Qwen3-Coder-30B-A3B-Instruct} & DFlash & 0.192 & 0.192 & 0.108 & 0.104 & $+43$ & $\mathbf{+46}$ \\
\code{gemma-4-26B-A4B-it} & MTP & 0.157 & 0.150 & 0.099 & 0.097 & $+38$ & $\mathbf{+35}$ \\
\code{JoyAI-LLM-Flash} & MTP & 0.247 & 0.182 & 0.159 & 0.121 & $+35$ & $\mathbf{+33}$ \\
\code{gemma-4-12B-it} & MTP & 0.139 & 0.134 & 0.105 & 0.098 & $+24$ & $\mathbf{+27}$ \\
\code{Qwen3-8B} & DSpark & 0.085 & 0.081 & 0.066 & 0.063 & $+23$ & $\mathbf{+22}$ \\
\code{Qwen3-14B} & DSpark & 0.137 & 0.136 & 0.111 & 0.110 & $+19$ & $\mathbf{+20}$ \\
\code{Qwen3-4B} & DSpark & 0.063 & 0.052 & 0.053 & 0.042 & $+16$ & $\mathbf{+19}$ \\
\code{GLM-4.7-Flash} & MTP & 0.229 & 0.180 & 0.161 & 0.153 & $+31$ & $\mathbf{+15}$ \\
\bottomrule
\end{tabular}

%% file: tab_tbt.tex
% GENERATED by analysis/collect_results.py (tpot_slo) -- do not edit.
\begin{tabular}{@{}lrrrrrr@{}}
\toprule
 & \multicolumn{3}{c}{baseline (ms)} & \multicolumn{3}{c}{fastest speculative scheme (ms)} \\
\cmidrule(lr){2-4}\cmidrule(lr){5-7}
model & $B{=}1$ & $B{=}32$ & $B{=}64$ & $B{=}1$ & $B{=}32$ & $B{=}64$ \\
\midrule
\code{GLM-4.7-Flash} & 8.2 & 23.4 & 28.0 & 6.2 {\scriptsize MTP} & 18.2 {\scriptsize MTP} & 21.5 {\scriptsize MTP} \\
\code{JoyAI-LLM-Flash}$^{\dagger}$ & 4.2 & 15.4 & 20.1 & 3.0 {\scriptsize MTP} & 10.8 {\scriptsize MTP} & 13.6 {\scriptsize MTP} \\
\code{Qwen3-14B} & 15.4 & 16.9 & 17.9 & 7.3 {\scriptsize DS} & 9.8 {\scriptsize DS} & 14.7 {\scriptsize DS} \\
\code{Qwen3-4B} & 5.7 & 6.8 & 7.7 & 2.7 {\scriptsize DS} & 4.3 {\scriptsize DS} & 6.0 {\scriptsize DS} \\
\code{Qwen3-8B} & 9.4 & 10.2 & 11.3 & 3.8 {\scriptsize DS} & 6.2 {\scriptsize DS} & 8.7 {\scriptsize DS} \\
\code{Qwen3-Coder-30B-A3B-Instruct} & 5.3 & 21.4 & 25.6 & 3.1 {\scriptsize DF} & 9.8 {\scriptsize DF} & 12.2 {\scriptsize DF} \\
\code{Qwen3.6-27B} & 15.1 & 20.3 & -- & 6.0 {\scriptsize DF} & 11.5 {\scriptsize MTP} & -- \\
\code{Qwen3.6-35B-A3B} & 4.4 & 11.2 & 14.5 & 3.4 {\scriptsize MTP} & 8.3 {\scriptsize MTP} & 11.4 {\scriptsize MTP} \\
\code{gemma-4-12B-it} & 13.9 & 16.0 & 18.1 & 3.5 {\scriptsize DF} & 9.3 {\scriptsize MTP} & 13.8 {\scriptsize MTP} \\
\code{gemma-4-26B-A4B-it} & 6.1 & 16.7 & 20.0 & 3.5 {\scriptsize DF} & 9.5 {\scriptsize MTP} & 13.5 {\scriptsize MTP} \\
\code{gemma-4-31B-it} & 33.2 & 38.3 & 56.6 & 7.8 {\scriptsize DF} & 19.0 {\scriptsize MTP} & 28.9 {\scriptsize MTP} \\
\bottomrule
\end{tabular}